\documentclass[conference]{IEEEtran}

\ifCLASSINFOpdf
  \usepackage[pdftex]{graphicx}
\else
  \usepackage[dvips]{graphicx}
\fi

\usepackage{amsmath,thmtools}
\usepackage{hyperref}
\usepackage{cases}
\usepackage{array}
\usepackage{color}
\usepackage[table]{xcolor}
\definecolor{mycolor}{rgb}{1,0,1} 

\newcommand*{\affaddr}[1]{#1} 
\newcommand*{\affmark}[1][*]{\textsuperscript{#1}}
\newcommand*{\email}[1]{\texttt{#1}}

\newcommand \ignore[1]{}

\begin{document}
%

\title{Implicit Authentication in Wearables \\Using Multiple Biometrics}
\title{Implicit Authentication of Wearable Device Users During Sedentary and Non-sedentary Periods}
\title{Sedentary and Non-sedentary Period \\Authentication of Wearable Device Users}
\title{Biometric-Based Wearable User Authentication During Sedentary and Non-sedentary Periods}

\author{%
Sudip Vhaduri\affmark[1], and Christian Poellabauer\affmark[1]\\
\affaddr{\affmark[1]Department of Computer Science and Engineering}\\
\affaddr{University of Notre Dame, IN 46556}\\
\email{\{svhaduri,cpoellab\}@nd.edu}\\
}


\maketitle

\begin{abstract}
The Internet of Things (IoT) is increasingly empowering people with an interconnected world of physical objects ranging from smart buildings to portable smart devices such as wearables. With the recent advances in mobile sensing, wearables have become a rich collection of portable sensors and are able to provide various types of services including health and fitness tracking, financial transactions, and unlocking smart locks and vehicles. 
Existing explicit authentication approaches (i.e., PINs or pattern locks) suffer from several limitations including limited display size, shoulder surfing, and recall burden. Oftentimes, users completely disable security features out of convenience. Therefore, there is a need for a burden-free (implicit) authentication mechanism for wearable device users based on easily obtainable biometric data.  
In this paper, we present an implicit wearable device user authentication mechanism using combinations of three types of coarse-grained minute-level biometrics: behavioral (step counts), physiological (heart rate), and hybrid (calorie burn and metabolic equivalent of task).
From our analysis of 421 Fitbit users from a two-year long health study, we are able to authenticate subjects with average accuracy values of around 92\% and 88\% during {\em sedentary} and {\em non-sedentary} periods, respectively. Our findings also show that (a) behavioral biometrics do not work well during {\em sedentary} periods and (b) hybrid biometrics typically perform better than other biometrics.
\end{abstract}


%
\IEEEpeerreviewmaketitle

\section{Introduction}\label{introduction}

With the rise of the Internet of Things (IoT), we are now able to remotely monitor and control physical objects, such as vehicles, buildings, health sensors, and many other smart devices. One specific example of such smart devices are wearables, with their ever improving sensing capabilities and network connectivity. Wrist-worn smart devices, such as fitness bands or smartwatches, are used for an increasing number of applications, including user 
identification for third party services~\cite{bianchi2016wearable}, creating a vault for sensitive information (i.e., passwords, credit card information)~\cite{nguyen2017smartwatches}, unlocking vehicles~\cite{nguyen2017smartwatches}, accessing phones and other paired devices\ignore{~\cite{kumar2016authenticating}}, managing financial payments~\cite{seneviratne2017survey}, health and fitness tracking, and monitoring of other individuals (e.g., child monitoring or fall detection of elderly people). 

While providing these new applications, wearables also introduce various new security and privacy challenges. For example, unauthorized access to a wearable device can provide an attacker with access to IoT systems controlled and monitored by the wearable~\cite{shahzad2017continuous,zeng2017wearia}. 
Wearables often also collect and store significant amounts of personal (and confidential) user data, which need to be protected from theft. As a consequence, it is essential to provide authentication and security mechanisms for these devices. Existing wearable device authentication mechanisms include knowledge-based regular PIN locks or pattern locks~\cite{nguyen2017smartwatches}, which suffer from scalability concerns~\cite{unar2014review}, since in the IoT world users are flooded with passwords/PINs to obtain access to various objects and services. 
Additionally, knowledge-based approaches require users to explicitly interact with a display (if present), which can be inconvenient to use~\cite{unar2014review,zeng2017wearia}. One consequence of this is that many users completely omit the authentication process and leave their devices vulnerable to attacks. Finally, knowledge-based approaches also suffer from observation attacks such as shoulder surfing~\cite{unar2014review}.
Therefore, in recent years, biometric-based solutions have been proposed, since they provide opportunities for implicit authentication, i.e., no direct user involvement or attention is required~\cite{unar2014review,zeng2017wearia}.
However, biometric-based authentication also has challenges and shortcomings, specifically in terms of accuracy and usability. For example, behavioral biometric-based approaches (e.g., gait and gesture) often fail to authenticate a user during periods of low physical activity (e.g., during sedentary tasks)~\cite{cola2016gait,zeng2017wearia}, and physiological biometric-based approaches (e.g., ECG or EEG signals) require very precise readings from expensive sensors, which are not available on most wearables due to computational and energy constraints~\cite{blasco2016survey}.
\ignore{, such as there is no need to remember anything, biometric attributes cannot be lost, transferred or stolen, offer better security due to fact that these attributes are very difficult to forge and require the presence of genuine user while granting access to particular resources}

\section{Related Work}\label{relatedwork}


Compared to mobile device user authentication, wearable device user authentication is a relatively new research area and traditional user authentication approaches are often not suitable for wearable devices, where computational capabilities and energy resources are much more constrained, or where low-cost sensors may be less accurate (noisy data recordings) or collect recordings only infrequently (e.g., once per minute)~\cite{blasco2016survey}. 
For example, most wearable health trackers make occasional heart rate measurements only instead of collecting raw and much more detailed (but also more costly in terms of energy and computational burden) ECG measurements. 
Recently researchers have proposed authentication techniques based on behavioral biometrics (e.g., gait~\cite{cola2016gait}\ignore{~\cite{al2017unobtrusive,cola2016gait,johnston2015smartwatch}}, gesture~\cite{davidson2016smartwatch}, and activity type~\cite{bianchi2016wearable,zeng2017wearia}) and physiological biometrics (e.g., PPG signals~\cite{karimian2017non}\ignore{~\cite{ohtsuki2016biometrie,sarkar2016biometric,karimian2017non}}). Almost all of these studies are based on\ignore{ fine-grained (i.e., multiple samples per second)} controlled data collections and the accuracy of these techniques has often been verified with limited numbers of subjects and over short time periods only. All of these user authentication techniques are also context dependent, e.g., behavioral biometric-based approaches do not work during {\em sedentary} periods, a model developed for one activity type does not work for other types, and heart rate values captured by a PPG sensor are affected by activity types and their intensities. Therefore, there is a need for a generic authentication approach that is able to consider different combinations of easily obtainable coarse-grained biometric data. 

\ignore{-------------------------------------------------------------------
User authentication for mobile devices has been an active research area for a long time~\cite{thullier2016exploring}. State-of-the-art techniques are based on various types of metrics, e.g., image~\cite{dantcheva2011bag,tisse2002person\ignore{,o2009context,balcan2005person}}, video~\cite{corvee2010person\ignore{,benabdelkader2002stride,benabdelkader2002person}}, voice~\cite{brunelli1995person\ignore{,hazen2003towards}}, acceleration and gait~\cite{kwapisz2010cell\ignore{,mantyjarvi2005identifying,ailisto2005identifying}}, Wi-Fi~\cite{zhang2016wifi}, Bluetooth low energy (BLE)~\cite{das2016uncovering}, electroencephalography (EEG) signals~\cite{poulos2002person,kumar2017bio}, electrocardiograms (ECG) signals~\cite{zhang2016review,choi2016biometric\ignore{,chan2008wavelet}}, and photoplethysmography (PPG) signals~\cite{sarkar2016biometric}. 
However, almost all of these traditional user authentication approaches are based on fine-grained sensor readings (i.e., many samples per second) from expensive sensors and hardware that have high computational capabilities and power capacities. Consequently, sensor readings are less noisy and provide person identification with a relatively high accuracy. 
Unfortunately, these traditional user authentication approaches are often not suitable for wearables, where computational capabilities and energy are much more constrained, or where low-cost sensors are less accurate (noisy data recordings) or collect recordings infrequently (e.g., once per minute)~\cite{blasco2016survey}. 
For example, most state-of-the-art health tracking wearables collect occasional heart rates instead of raw (and detailed) ECG measurements~\cite{choi2016biometric}.
Additionally, the lack of traditional input devices (e.g., keypad) and output devices (e.g., display) provide additional constraints on the choice of authentication mechanism~\cite{bianchi2016wearable}.

However, while the accuracy and sampling rates of modern wearables are limited, we claim that these measurements can still provide enough information to capture activity and fitness characteristics of an individual, thereby we can use the measurements to identify if a device is worn by its owner or another user. Another advantage of an authentication approach using physiological and activity metrics is that such user identification can run \textcolor{red}{\bf continuous}ly, thereby providing non-stop authentication. 

-------------------------------------------------------------------}
\section{Approach}

In this work, we propose an implicit and reliable wearable device user authentication scheme that relies on coarse-grained minute-level biometrics that are widely available on state-of-the-art wearables. While the combination of multiple biometrics will result in highly accurate user identification, the reliance on coarse-grained readings from sensors that are commonly found on most fitness and health trackers makes the proposed solution easy to deploy and resource efficient.
Compared to our previous work~\cite{vhaduri2017towards,vhaduri2017wearable}, in this paper, we investigate how different combinations of four common biometrics perform when authenticating users during both {\em sedentary} and {\em non-sedentary} periods.
Before we describe the details of the authentication models, we first discuss the dataset, pre-processing steps, feature computation, and feature selection.
For the following analysis we use minute-level {\bf heart rate}, {\bf calorie burn}, {\bf step counts}, and {\bf metabolic equivalent of task (MET)} as sensor data.

\subsection{NetHealth Study Dataset}\label{study}
The {\em NetHealth} mobile crowd sensing (MCS) study~\cite{vhaduri2017towards,vhaduri2017wearable,vhaduri2018hierarchical,vhaduri2018opportunisticTBD,vhaduri2018impact,vhaduri2016assessing,vhaduri2016human,vhaduri2016cooperative,vhaduri2017design,vhaduri2016design,vhaduri2018opportunisticICHI,vhaduri2017discovering} began at the University of Notre Dame in 2015\ignore{ with the purpose of investigating the impacts of ``always-on connectivity'' on the health habits, emotional wellness, and social ties of college students}.
For this study, over 400 individuals\ignore{subjects} were recruited from the freshmen class and the students were instructed to continuously wear a Fitbit Charge HR device that was provided to them.
%
The data being collected by the Fitbit devices include minute-level heart rate, average heart rate\ignore{~\cite{hr}}, calorie burn\ignore{~\cite{calorie}}, metabolic equivalent of task or MET\ignore{~\cite{met}}, physical activity level/intensity (e.g., sedentary, light, fair, and high)\ignore{~\cite{activity}}, step count, sleep status, and self-recorded activity labels. These collected data can be divided into three biometric groups: {\bf behavioral} (e.g., step counts, activity level/intensity), {\bf physiological} (e.g., heart rate), and {\bf hybrid} (e.g., calorie burn, MET) biometrics, where hybrid biometrics are derived from both behavioral and physiological biometrics.

\subsection{Data Pre-Processing and Feature Computation}\label{features}

Since we are using a real-world dataset, we first need to clean the dataset before using it. Then, we need to segment the continuous stream of biometrics, followed by feature computations before we can build our authentication models.

\subsubsection{Filtering Invalid Activity Data}

A Fitbit device collects heart rate data only when the device is actually worn, but the device collects activity data all the time, even if the device is not worn. Therefore, before we can use the activity data for our analysis, we need to remove ``invalid'' periods\ignore{ of activity minutes that do not match with periods of heart rate measurements}, i.e., the device is not worn.
For our analysis, we consider data from 421 Fitbit users.

\subsubsection{Data Segmentation and Feature Computation}

For the classification task, we first segment continuous heart rate, calorie burn, MET, and step counts into five-minute non-overlapping windows starting from a change of activity levels. Since the sampling rate is one sample per minute, each window contains five consecutive samples. When we segment the data into windows, we start from the beginning of an activity level and check for the next five minutes if the same activity level continues. With this approach, we set the reference point at the beginning of an activity level, since the biometrics vary across different activity levels\ignore{ as well as at different parts of the same activity level (e.g., beginning versus middle)}.

For each biometric, we compute 31 statistical features in both time and frequency domains: mean ($\mu$), standard deviation ($\sigma$), variance ($\sigma^2$), coefficient of variation ($cov$), maximum ($max$), minimum ($min$), range ($ran$), coefficient of range ($coran$), percentiles ($p25$, $p50$, $p75$, and $p95$), inter quartile range ($iqr$), coefficient of inter quartile range ($coi$), mean absolute deviation ($mad\_\mu$), median absolute deviation ($mad\_Mdn$), mean frequency ($f\_\mu$), median frequency ($f\_Mdn$), power ($P$), number of peaks ($np$), energy ($E$), root mean square ($rms$), peak magnitude to rms ratio ($p2rms$), root sum of squares ($rss$), signal to noise ratio ($snr$), skewness ($\gamma$), kurtosis ($\kappa$),
amplitude of the main frequency ($a\_main$) and secondary frequency ($a\_sec$), and main frequency ($f\_main$) and secondary frequency ($f\_sec$) of the {\em Discrete Fourier Transform} (DFT) signal obtained using the {\em Fast Fourier Transform} (FFT)\ignore{~\cite{wang1984fast}} function for each window of biometric data.
For {\em non-sedentary} periods, we also consider the activity level as an additional feature. Therefore, we compute a maximum of 124 and 125 features for each window during {\em sedentary} and {\em non-sedentary} periods, respectively.

In the rest of this paper, each biometric is referred to by its initial: ``C'' (calorie burn), ``S'' (step count), ``M'' (MET), and ``H'' (heart rate). Combinations of these letters are used to represent the corresponding combinations of the biometrics, e.g., ``CH'' represents a combination of calorie burn and heart rate. Therefore, a biometric combination $b \in$ \{C, S, M, H, CS, CM, CH, SM, SH, MH, CSM, CSH, CMH, SMH, CSMH\}.

\subsection{Feature Selection}\label{feature_selection}
To find relevant features, we first use the {\em Two-sample Kolmogorov-Smirnov} (KS)-test\ignore{~\cite{smirnov1948table}} with the null hypothesis $H_0$: ``the two data sets are from the same distribution.'' For each feature, we calculate the $p$-$value$ for data points from each pair of subjects and drop a feature if most of its $p$-$values$ are higher than $\alpha = .05$, i.e., the non-discriminating features. We find that during {\em sedentary} minutes the behavioral biometric (step count) has no significant feature. However, the behavioral biometric contributes to a good number of significant features during {\em non-sedentary} (i.e., {\em lightly}, {\em fairly}, and {\em highly} active) minutes.


Next, we apply the Coefficient of Variation (COV)-approach on features obtained from the KS-test. The feature that varies more (i.e., higher $cov$ values) across subjects has a higher chance of capturing subject varying information, i.e., it can be an influential feature and can better distinguish the subject compared to less influential features that do not vary much. Compared to our previous standard deviation-based approach, the COV-approach is a better measure when comparing different features since $cov$ is a measure of relative variability, i.e., $cov = \sigma/\mu$.
For each biometric combination and its associated feature set, we compute the $cov$ of all features in the set and then we find the maximum of the $cov$ values of all features in the set. Next, we compute a set of thresholds using $x^{\sigma t} \in \{10, 20, ..., 90\}$ percent of that maximum $cov$ value. Finally, for each threshold, we pick only those features that have $cov$ values higher than the threshold.

Finding a proper threshold $x^{\sigma t}$ can be tricky; if it is chosen too small, this may lead to a feature set containing redundant and less important features, which may lead to overfitting. In contrast, if the threshold is chosen too high, this may lead to a very small feature set and poor accuracy.
In Section~\ref{param_opt}, we present the optimal values of $x^{\sigma t}$.

A sample feature set obtained using the COV-approach during {\em non-sedentary} periods with $b = CM$ and $x^{\sigma t} = 30$\% consists of $27$ features: ``C'' ($\mu$, $\sigma$, $max$, $min$, $ran$, $p25$, $p50$, $p75$, $p95$, $iqr$, $mad\_\mu$, $mad\_Mdn$, $rms$, $rss$, $a\_main$, $a\_sec$) and ``M'' ($\mu$, $max$, $p25$, $p50$, $p75$, $p95$, $P$, $E$, $rms$, $rss$, $a\_main$).

\section{User Authentication}
\label{authentication}

In this section, we analyze the performance of different feature sets using the binary {\em Quadratic Support Vector Machine} (q-svm), the best classifier 
to authenticate wearable device users as found in our previous work~\cite{vhaduri2017wearable}. 
Before analyzing the performance, we prepare our training-testing datasets. When preparing the datasets, the number of windows that we consider is at least 10 times the number of features in the set. This helps to avoid overfitting. For each feature set, we further balance the dataset by randomly selecting the same number of windows per activity level per subject.
Next, we split the entire dataset into 75\%--25\% for training and testing.

\begin{figure}[!t]
\centering
\includegraphics[width=3.4in]{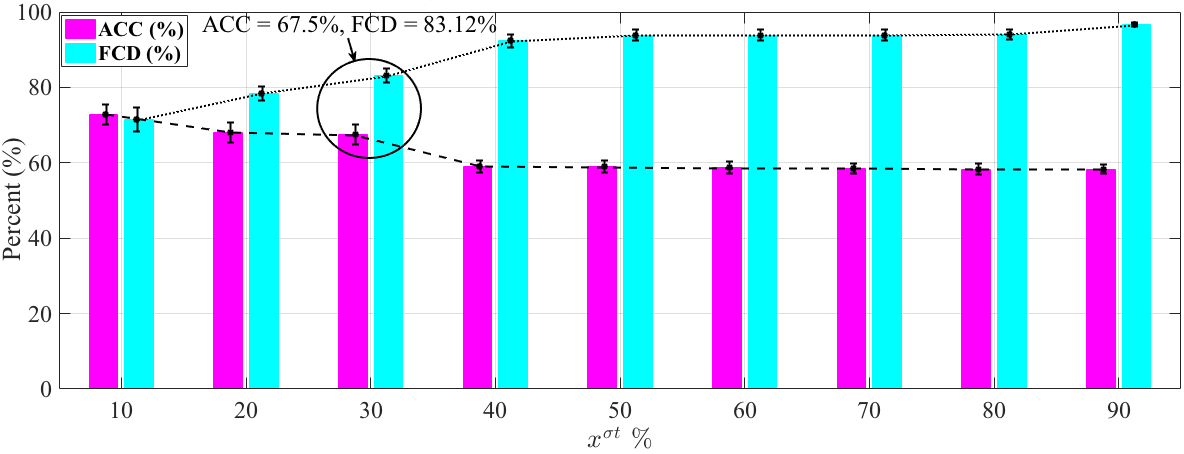}
\caption{Average ACC and FCD across different values of parameter $x^{\sigma t}$ during {\em non-sedentary} periods.}
\label{param_opti_sed_nonSed}
\end{figure}

\subsection{Performance Measures}

To evaluate the performance of different feature sets we use {\em Accuracy} (ACC) (in \%) as the primary measure, which is the fraction of predictions that are correct, i.e., $(TP+TN)/(TP+TN+FP+FN)\times 100\%$, where terminologies have their usual meaning in machine learning.
We also consider {\em Feature Count Decrease} (FCD) (in \%) as an additional performance measure. This is a measure of improvement in feature count that a feature selection approach can achieve defined as $FCD = (n^T-n)/n^T\times 100\%$, where $n^T$ and $n$ are the maximum number of features in the initial feature set that we start with (i.e., $124$ for {\em sedentary} and $125$ for {\em non-sedentary} periods with $b = CSMH$) and the number of features in a feature set, respectively. If two feature sets achieve the same accuracy, then the set with higher FCD, i.e., lower feature count, is better since it will lower the computational load, while achieving the same accuracy as the other set.

\subsection{User Authentication Models}

When building authentication models for a feature set with $N$ subjects (each having $|W|$ random windows), we train and test $N$ binary q-svm classifiers. Each of these $N$ classification models is used to authenticate a subject from the other $N-1$ subjects. 
Each subject is identified by an anonymous subject ID.
We perform wearable device user authentication separately for {\em sedentary} and {\em non-sedentary} periods. First, we find the optimal sets of parameters for different feature selection approaches (Section~\ref{param_opt}). Next, for each feature selection approach, using its optimal parameter set, we then compare the performance of different biometrics to find the best biometric combination (Section~\ref{best_bio_per_selec_app}).

\begin{figure}[!t]
\centering
\includegraphics[width=3.4in]{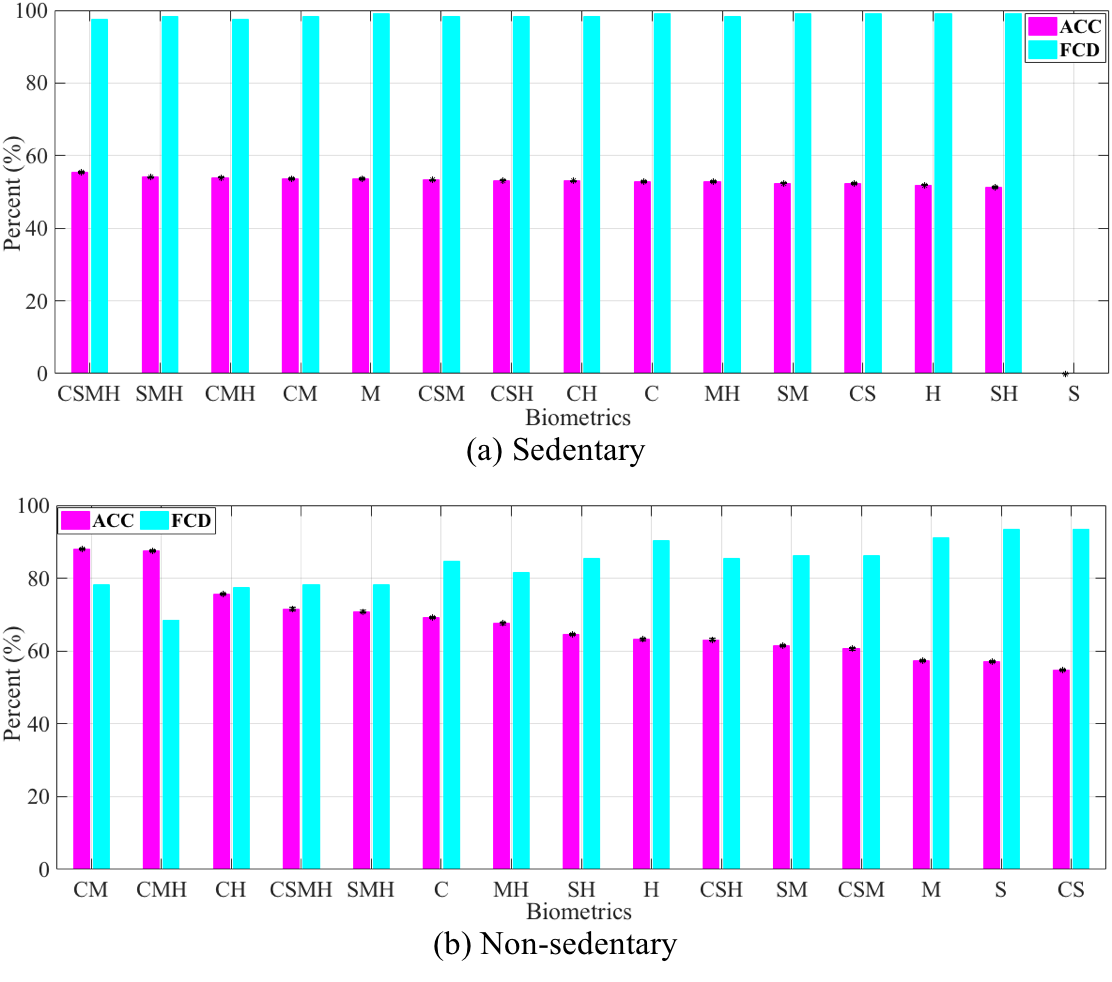} 
\caption{Bar graphs of authentication Average ACC (in \%) and FCD (in \%) variations across different biometrics of the COV-approach. The bars inside each subplot are sorted based on average ACC and FCD values. The best biometrics obtained from the two subplots are (a) $b = CSMH$ and (b) $b = CM$.}
\label{acc_feat_sets_sed_nonSed}
\end{figure} 

\subsubsection{Finding Optimal Parameter Sets}\label{param_opt}

To find the optimal parameter set for each feature selection approach, we first compute the average of all ACC and FCD values obtained for all possible combinations of subjects and biometrics. Then graphically we determine the optimal parameter setting. 
Figure~\ref{param_opti_sed_nonSed} shows an example of optimal parameter selection for the COV-approach during {\em non-sedentary} periods. 
In this figure we observe that with the increase of $x^{\sigma t}$ the average ACC decreases, but FCD increases. Therefore, we try to find an optimal value of $x^{\sigma t}$ at which both ACC and FCD achieve higher values. We pick $x^{\sigma t} = 30$\% as our optimal value since after this $x^{\sigma t}$ ACC drops and reaches saturation. Similarly, FCD reaches saturation after $x^{\sigma t} = 30$\%  (Figure~\ref{param_opti_sed_nonSed}).   
We obtain the $x^{\sigma t}$ threshold value for {\em sedentary} periods using the same approach. 

\begin{table}[!t]
\caption{Authentication Accuracy Summary ($l = 0$ is {\em sedentary} and $l = 1$ is {\em non-sedentary}}
\label{table:acc_summ_feat_sets}
\centering
\begin{tabular}{c|l|c|c|c}
\hline
$l$ & App-       & mean (SD)  & mean (SD) & Best biometric's mean           \\
    & -roach     & ACC        & FCD       & ACC ($b,n,N,|W|$)  \\  
\hline
0 & COV & 53.12 (1.03) & 98.62 (0.59) & 55.46 (CMH,3,415,475) \\
 & KS & 76.26 (12.46) & 64.06 (15.24) & 91.71 (CM,53,412,544)  \\
\hline
1 & COV & 68.24 (10.03) & 83.24 (6.67) & 88.00 (CM,27,332,331) \\
 & KS & 73.89 (9.80) & 70.97 (13.26) & 88.40 (CM,30,332,331)  \\
\hline
\end{tabular}
\end{table}

\subsubsection{Comparing Biometrics of Each Feature Selection Approach}\label{best_bio_per_selec_app}

First, we investigate how classifier performance varies across different biometrics for the same feature selection approach. Figure~\ref{acc_feat_sets_sed_nonSed} shows the ACC and FCD variation across different biometrics and their associated feature sets obtained from the COV-approach.
In Figure~\ref{acc_feat_sets_sed_nonSed} (a) we observe that during {\em sedentary} periods all biometrics except the behavioral biometric (i.e., step counts) perform similarly.\ignore{ Therefore, all existing approaches in the literature that rely on behavioral biometrics will fail to authenticate users during {\em sedentary} periods, which we can successfully tackle using our multi-modal biomteric-based authentication approach.} During {\em non-sedentary} periods $b = CM$ has the best performance compared to the other 14 biometrics (Figure~\ref{acc_feat_sets_sed_nonSed} (b)). 
Table~\ref{table:acc_summ_feat_sets} summarizes the user authentication performance, where the average ACC and FCD values are computed from all possible 15 biometric combinations under a specific feature selection approach. Similarly, the last column in the table also represents an average ACC, but it is computed for a particular biometric combination under a specific feature selection approach. For example, we obtain an average ACC = $55.46$ for $b = CMH$ under the COV-approach during {\em sedentary} periods. On average the KS-approach achieves a better ACC compared to the COV-approach. However, the KS-approach has a poor average FCD compared to the COV-approach.
In the last column (i.e., ``Best biometric's mean ACC'' column) in Table~\ref{table:acc_summ_feat_sets} we observe that the two hybrid biometrics (calorie burn (C) and MET (M)) together perform better than other biometrics. During {\em non-sedentary} periods the KS- and COV-approaches have similar performances. However, during {\em sedentary} periods there is a big difference between KS- and COV-approaches.

\section{Conclusions\ignore{ and Future Work}}
\label{conclusions}

To our best knowledge, our work is the first to use three different types of less informative coarse-grained processed biometric data (i.e., behavioral, physiological, and hybrid) to authenticate the wearable device users implicitly during both {\em sedentary} and {\em non-sedentary} periods. 

Our findings from the different combinations of the four biometrics (Section~\ref{best_bio_per_selec_app}) show that when behavioral biometrics (step counts) fail to authenticate a user during {\em sedentary} periods, our multi-modal biometric-based approach can still authenticate the users with a good average accuracy (around 92\% with {\em Genuine Acceptance Rate (GAR)} = .98\ignore{ and {\em False Acceptance Rate (FAR)} = .13}, obtained from a set of 412 subjects). Similarly, for {\em non-sedentary} periods we achieve an average accuracy of 88\% with {\em GAR} = .99\ignore{ and {\em FAR} = .22} using only $27$ features (based on a set of 332 subjects). 
In general, we find that the hybrid biometrics (calorie burn and MET) achieve better performance compared to other biometrics. 
These accuracy values can further be improved by considering various spatio-temporal factors that can impact person-dependent biometrics. 
However, to make the authentication approach generic, we build models with relatively smaller feature sets.

\bibliographystyle{IEEEtran}
\bibliography{reference}

\begin{thebibliography}{10}
\providecommand{\url}[1]{#1}
\csname url@samestyle\endcsname
\providecommand{\newblock}{\relax}
\providecommand{\bibinfo}[2]{#2}
\providecommand{\BIBentrySTDinterwordspacing}{\spaceskip=0pt\relax}
\providecommand{\BIBentryALTinterwordstretchfactor}{4}
\providecommand{\BIBentryALTinterwordspacing}{\spaceskip=\fontdimen2\font plus
\BIBentryALTinterwordstretchfactor\fontdimen3\font minus
  \fontdimen4\font\relax}
\providecommand{\BIBforeignlanguage}[2]{{%
\expandafter\ifx\csname l@#1\endcsname\relax
\typeout{** WARNING: IEEEtran.bst: No hyphenation pattern has been}%
\typeout{** loaded for the language `#1'. Using the pattern for}%
\typeout{** the default language instead.}%
\else
\language=\csname l@#1\endcsname
\fi
#2}}
\providecommand{\BIBdecl}{\relax}
\BIBdecl

\bibitem{bianchi2016wearable}
A.~Bianchi and I.~Oakley, ``Wearable authentication: Trends and
  opportunities,'' \emph{it-Information Technology}, vol.~58, no.~5, pp.
  255--262, 2016.

\bibitem{nguyen2017smartwatches}
T.~Nguyen and N.~Memon, ``Smartwatches locking methods: A comparative study,''
  in \emph{Symposium on Usable Privacy and Security}, 2017.

\bibitem{seneviratne2017survey}
S.~Seneviratne, Y.~Hu, T.~Nguyen, G.~Lan \emph{et~al.}, ``A survey of wearable
  devices and challenges,'' \emph{IEEE Communications Surveys \& Tutorials},
  vol.~19, no.~4, pp. 2573--2620, 2017.

\bibitem{shahzad2017continuous}
M.~Shahzad and M.~P. Singh, ``Continuous authentication and authorization for
  the internet of things,'' \emph{IEEE Internet Computing}, 2017.

\bibitem{zeng2017wearia}
Y.~Zeng, A.~Pande, J.~Zhu, and P.~Mohapatra, ``Wearia: Wearable device implicit
  authentication based on activity information,'' in \emph{IEEE World of
  Wireless, Mobile and Multimedia Networks (WoWMoM)}, 2017.

\bibitem{unar2014review}
J.~Unar, W.~C. Seng, and A.~Abbasi, ``A review of biometric technology along
  with trends and prospects,'' \emph{Pattern recognition}, vol.~47, no.~8, pp.
  2673--2688, 2014.

\bibitem{cola2016gait}
G.~Cola, M.~Avvenuti, F.~Musso, and A.~Vecchio, ``Gait-based authentication
  using a wrist-worn device,'' in \emph{Proc. Mobile and Ubiquitous Systems:
  Computing, Networking and Services}.\hskip 1em plus 0.5em minus 0.4em\relax
  ACM, 2016.

\bibitem{blasco2016survey}
J.~Blasco, T.~M. Chen, J.~Tapiador, and P.~Peris-Lopez, ``A survey of wearable
  biometric recognition systems,'' \emph{ACM Computing Surveys (CSUR)},
  vol.~49, no.~3, p.~43, 2016.

\bibitem{davidson2016smartwatch}
S.~Davidson, D.~Smith, C.~Yang, and S.~Cheah, ``Smartwatch user identification
  as a means of authentication,'' \emph{Department of Computer Science and
  Engineering Std}, 2016.

\bibitem{karimian2017non}
N.~Karimian, M.~Tehranipoor, and D.~Forte, ``Non-fiducial ppg-based
  authentication for healthcare application,'' in \emph{IEEE BHI}, 2017.

\bibitem{vhaduri2017towards}
S.~Vhaduri and C.~Poellabauer, ``Towards reliable wearable-user
  identification,'' in \emph{IEEE ICHI}, 2017.

\bibitem{vhaduri2017wearable}
------, ``Wearable device user authentication using physiological and
  behavioral metrics,'' in \emph{IEEE PIMRC}, 2017.

\bibitem{vhaduri2018hierarchical}
------, ``Hierarchical cooperative discovery of personal places from location
  traces,'' \emph{IEEE Transactions on Mobile Computing}, 2018.

\bibitem{vhaduri2018opportunisticTBD}
------, ``Opportunistic discovery of personal places using multi-source sensor
  data,'' \emph{IEEE Transactions on Big Data}, 2018.

\bibitem{vhaduri2018impact}
------, ``Impact of different pre-sleep phone use patterns on sleep quality,''
  in \emph{IEEE Wearable and Implantable Body Sensor Networks (BSN)}, 2018.

\bibitem{vhaduri2016assessing}
S.~Vhaduri, A.~Munch, and C.~Poellabauer, ``Assessing health trends of college
  students using smartphones,'' in \emph{IEEE HI-POCT}, 2016.

\bibitem{vhaduri2016human}
S.~Vhaduri and C.~Poellabauer, ``Human factors in the design of longitudinal
  smartphone-based wellness surveys,'' in \emph{IEEE ICHI}, 2016.

\bibitem{vhaduri2016cooperative}
------, ``Cooperative discovery of personal places from location traces,'' in
  \emph{ICCCN}, 2016.

\bibitem{vhaduri2017design}
------, ``Design factors of longitudinal smartphone-based health surveys,''
  \emph{Journal of Healthcare Informatics Research}, 2017.

\bibitem{vhaduri2016design}
------, ``Design and implementation of a remotely configurable and manageable
  well-being study,'' in \emph{Smart City 360°}, 2016.

\bibitem{vhaduri2018opportunisticICHI}
------, ``Opportunistic discovery of personal places using smartphone and
  fitness tracker data,'' in \emph{IEEE ICHI}, 2018.

\bibitem{vhaduri2017discovering}
S.~Vhaduri, C.~Poellabauer \emph{et~al.}, ``Discovering places of interest
  using sensor data from smartphones and wearables.''

\end{thebibliography}

\end{document}